\documentclass[10pt,conference]{IEEEtran}
\IEEEoverridecommandlockouts
\usepackage{cite}
\usepackage{amsmath,amssymb,amsfonts}
\usepackage{algorithmic}
\usepackage{graphicx}
\usepackage{textcomp}
\usepackage{xcolor}
\def\BibTeX{{\rm B\kern-.05em{\sc i\kern-.025em b}\kern-.08em
    T\kern-.1667em\lower.7ex\hbox{E}\kern-.125emX}}

\usepackage{graphicx}
\usepackage{float}
\usepackage{subcaption}
\usepackage{caption}
\usepackage[most]{tcolorbox}
\usepackage{dashrule}
\usepackage[dvipsnames]{xcolor}
\usepackage{adjustbox}
\usepackage{bm}
\usepackage{multirow}
\usepackage{adjustbox}
\usepackage{booktabs}
\usepackage{makecell}
\usepackage{tabularx}
\usepackage{array}
\usepackage{enumitem}
\usepackage{hyperref}
\usepackage{balance}

\newcommand{\revised}[1]{\textcolor{black}{#1}}

\begin{document}

\title{COMPASS: Predicting the Relationship of Multiple Patches for Vulnerabilities with LLMs}

\author{
\IEEEauthorblockN{
Yi Song\textsuperscript{1,*},
Dongchen Xie\textsuperscript{1,*},
Xiaoyuan Xie\textsuperscript{2,$\dagger$},
He Zhang\textsuperscript{2},
Lin Xu\textsuperscript{1},
Chunying Zhou\textsuperscript{2},
and Zhi Jin\textsuperscript{2}
}
\IEEEauthorblockA{
\textsuperscript{1}\textit{School of Cyber Science and Engineering, Wuhan University, Wuhan, China}\\
\textsuperscript{2}\textit{School of Computer Science, Wuhan University, Wuhan, China}\\
\{yisong, xiedongchen, xxie, zhanghe, xulin\_xl, zcy9838, zhijin\}@whu.edu.cn
}
}

\maketitle

\begingroup
\renewcommand{\thefootnote}{\fnsymbol{footnote}}
\footnotetext[1]{Yi Song and Dongchen Xie contributed equally to this work.}
\footnotetext[2]{Xiaoyuan Xie is the corresponding author.}
\endgroup

\begin{abstract}
Modern software heavily relies on code reuse, so upstream vulnerability fixes do not automatically propagate to downstream codebases. Downstream maintainers must manually adopt patches to eliminate known risks. In practice, a single vulnerability often corresponds to multiple patches, which greatly complicates downstream patch adoption because different patch relationships imply different adoption strategies. To address this challenge, \revised{we first manually inspect large-scale multi-patch vulnerabilities ($\sim$1K) in the real world and interview experienced developers}, summarizing six typical types of patch relationships, i.e., \emph{Merge}, \emph{Mirror}, \emph{Better Solution}, \emph{Fixing-of-Fixing}, \emph{Collaboration}, and \emph{Separation}. Based on these observations, we propose COMPASS, an automated approach that predicts the relationships of multiple vulnerability patches with large language models. Given a CVE as input, COMPASS follows a four-phase pipeline that (\emph{i}) \revised{identifies the patch group and pre-scans explicit relationships}, (\emph{ii}) performs individual patch analysis, (\emph{iii}) infers relationship instances via a hierarchy-guided prompt, and (\emph{iv}) validates completeness and consistency of the inferred results. As output, COMPASS reports the predicted relationships within the patch group and visualizes them as a relationship graph. We evaluate COMPASS on a benchmark of 300 multi-patch CVEs and compare it against mainstream learning-based \revised{and LLM} baselines. \revised{Results show that our method achieves strong and consistent prediction effectiveness and outperforms SOTA by 85.04\% on average.} We publicly release an online querying website to support community reuse of patch relationships knowledge: \href{https://patch-relation.com}{https://patch-relation.com}.
\end{abstract}

\begin{IEEEkeywords}
Security Patches, Patch Relationships, Open-source Community, Large Language Models
\end{IEEEkeywords}

\section{Introduction}
\label{sect:introduction}

Security vulnerabilities pose severe risks to software systems, patches (typically in the form of code commits\footnote{We use the terms ``\emph{commit}'' and ``\emph{patch}'' interchangeably unless otherwise specified, because the commits we are focusing on to fix vulnerabilities are actually patches.}) are the most important and straightforward resources to remediate their hazard~\cite{9152613,10.1145/3749370,10.1145/3694782}. In practice, one vulnerability is often fixed by more than one patch rather than a single commit, i.e., the ``\emph{many-for-one}'' fixing pattern. Existing research has shown that this phenomenon is widespread in real-world software ecosystems. For example, Xu et al. reveal that multiple patches are developed for about 41\% of the vulnerabilities by an empirical study~\cite{xu2022tracking}, Jiang et al. point out that one vulnerability may map to more than one patch due to incomplete fixes~\cite{jiang2024understanding}, and Li et al. find that a non-trivial portion of vulnerability entries on public platforms such as CVE (Common Vulnerabilities and Exposures)~\cite{cve-website} and NVD (National Vulnerability Database)~\cite{nvd-website} has multiple associated code commits. Apart from these, there are many other studies demonstrating the prevalence of this many-for-one fixing pattern~\cite{li2017large,tan2021locating, xu2022tracking,hommersom2024automated,woo2025large}.

In the context of open-source environments, the general many-for-one fixing pattern in upstream software brings significant complexity to patch adoption for downstream projects that depend on it. The complexity is largely determined by the relationships of patches, as different relationships directly lead to different adoption strategies. Specifically, modern software development heavily relies on code reuse: many projects act as upstream software platforms whose code is copied, integrated, and customized by a large number of downstream projects, and upstream vulnerability fixes do not propagate to downstream code by default because the code between the upstream and downstream is disconnected~\cite{10.1007/s10664-021-09951-x,10.1007/978-3-032-05188-2_21,11025716,Zhang2021AnIO,Haq2025TheRE}. Thus, when an upstream vulnerability is fixed, downstream projects need to manually adopt these patches. However, the decoupling caused by code copying, long-term independent evolution, and project-specific customization often makes upstream patches inapplicable or ambiguous, and more importantly, the many-for-one fixing pattern implies that these patches may play different roles and interact in complex ways. As a result, downstream maintainers must understand how these patches relate to each other (i.e., their relationships) before deciding how to adopt them~\cite{10555642,10.1145/3460319.3464821,Fedora,li2025fixllmaidedcategorizationsecurity}.

For example, if Patch-$A$ initially fixes a vulnerability but inadvertently introduces a new bug or vulnerability, and Patch-$B$ later fixes this issue, the two patches must be applied sequentially and released together, otherwise, releasing the two patches in different releases (e.g., due to the tight schedule, the capability of the maintenance team, or the limitation of available manpower) may expose users to new security risks. In another case, multiple patches may collaboratively fix a vulnerability step by step, where later patches just depend on variables or logic introduced by earlier ones. Such patches must be applied in order, but they can be released separately when necessary, because intermediate patches represent incomplete yet safe fixes that do not introduce new flaws. For another example, if Patch-$A$ provides a complete fix and Patch-$B$ merely optimizes the original solution (e.g., improving performance or code readability), downstream maintainers may choose to apply only the latter to avoid redundant work. In addition to the above example situations, there are more other cases where the relationship of patches largely participates in the vulnerability fixing, which will be given in Section~\ref{sect:typicalrelationshipsofpatches}. These examples illustrate that different patch relationships imply fundamentally different adoption behaviors, and misunderstanding these relationships can lead to delayed or inconsistent decisions, and a higher risk of incomplete or even unsafe patch adoption.

Therefore, a precise understanding of patch relationships is essential for facilitating effective patch adoption in downstream software, which directly impacts the quality and security of end-user software applications. However, such understanding is currently obtained almost exclusively through manual analysis, which is not only time-consuming and labor-intensive, but also inherently error-prone and difficult to scale~\cite{xu2022tracking,10.1145/2568225.2568260,Ogenrwot2025RefactoringAwarePI}. Maintainers need to carefully inspect commit messages, examine code changes across multiple patches, and reason about subtle inter-patch dependencies. As the number of vulnerabilities and associated patches continues to grow, this manual process becomes increasingly impractical.

To address this challenge, we propose COMPASS, an automated approach of predi\textbf{C}ting the relati\textbf{O}nship of \textbf{M}ulti\textbf{P}le p\textbf{A}tche\textbf{S} for vulnerabilitie\textbf{S} with large language models (LLMs). \revised{As a preliminary, we inspect large-scale vulnerability cases (1,050 CVE entries) having multiple patches in the wild, and interview 20 experienced developers,} summarizing six typical types of relationships of patches, i.e., \emph{Merge}, \emph{Mirror}, \emph{Better Solution}, \emph{Fixing-of-Fixing}, \emph{Collaboration}, and \emph{Separation} (which will be explained in detail in Section~\ref{sect:typicalrelationshipsofpatches}). Driven by these typical relationships, COMPASS performs a fourfold pipeline to predict patches' relationships. First, for a given vulnerability, we determine all its associated patches (referred to as \emph{patch group}) by checking the corresponding \revised{NVD} webpage for recorded patch links, \revised{and pre-scanning explicit relationships}. Second, we feed the patch group (including the commit ID, the branch/version information, the commit message, and the code change of each patch) to LLMs, and guide LLMs to think about three questions using the intuition of chain-of-thought: (\emph{i}) double-checking whether each patch \revised{directly} fixes the vulnerability, (\emph{ii}) identifying key components/files modified by the patches, and (\emph{iii}) noting implicit signals available in commit messages and branch/version information. Third, we design a sophisticated prompt describing the summarized six relationships in the text form, and ask LLMs to predict the relationships of the fed patches according to given schema based on the thoughts. Fourth, LLMs are required to validate the prediction result to make sure that all patches have been predicted, no contradictory relationships, and so on. Finally, the relationships \revised{within the patch group} will be output and further reorganized into a visualized graph to facilitate users' understanding.

To evaluate the effectiveness of COMPASS, we construct a benchmark consisting of 300 vulnerability entries with 2 $\sim$ 7 patches, \revised{which is completely disjoint from the 1,050 multi-patch vulnerability entries used for taxonomy construction.} The benchmark entries are collected from public platforms CVE~\cite{cve-website}, NVD~\cite{nvd-website}, and Snyk~\cite{snyk-website}, span the period from 2009 to 2024, and each entry includes a CVE identifier, all associated patches and their relationships. We further develop an online website to publicly host vulnerability entries together with the relationships of their multiple patches. This design is motivated by the observation that a single CVE can affect many downstream projects, and thus the relationships of its patches are often repeatedly queried by different teams, making a shared resource valuable for avoiding duplicated manual effort. The 300 benchmark entries have been uploaded as an initial release, and we will continuously maintain the website in the future. We also implement COMPASS as a user-friendly tool that, \revised{given the identifier of a CVE having multiple patches, automatically predicts its patch relationships, and visualizes the results to support practical downstream patch adoption.}

The main contributions of this paper are as follows:

\begin{itemize}
    \item \textbf{A novel approach.} We propose COMPASS to automatically predict the relationship of multiple patches for a given vulnerability, for facilitating downstream maintainers' understanding for better adoption. To the best of our knowledge, this is the first work focusing on vulnerability patches' relationship prediction.

    \item \textbf{A manually-crafted benchmark.} We construct a benchmark of 300 real-world vulnerabilities with multiple patches, where patch relationships are manually analyzed and labeled for evaluation. We make this dataset publicly available \revised{to facilitate future research}.

    \item \textbf{An online querying website and a tool.} We develop and maintain a public website to host patch relationships for community reuse. We also implement COMPASS as a user-friendly tool to automatically predict and visualize patches' relationships of disclosed vulnerabilities.

\end{itemize}

The experimental code (including the designed prompt) and data, as well as the source code and the instructions for the tool, are given in a public repository~\cite{publicrepository}. The URL of the online querying website is \urlstyle{rm}{\textbf{\url{https://patch-relation.com}}}.

The remainder of this paper is organized as follows: Section~\ref{sect:typicalrelationshipsofpatches} describes the summarized six typical types of patch relationships. Section~\ref{sect:approach} describes the details of COMPASS. Section~\ref{sect:experimentalsetup} provides the research questions, baselines, datasets, and metrics. Section~\ref{sect:resultandanalysis} analyzes the experimental results. Section~\ref{sect:discussion} further discusses our approach. Section~\ref{sect:threatstovalidity} declares the threats to
validity. Section~\ref{sect:relatedwork} summarizes the related work. Section~\ref{sect:conclusion} concludes the paper and outlines future work.

\section{Typical Relationships of Patches}
\label{sect:typicalrelationshipsofpatches}

\subsection{Taxonomy Construction}
\label{subsect:taxonomyconstruction}

\revised{To describe how multiple patches associated with the same vulnerability relate to each other in the real world, we preliminarily inspected 1,050 multi-patch vulnerability entries from CVE~\cite{cve-website}, NVD~\cite{nvd-website}, and Snyk~\cite{snyk-website}. We constructed the taxonomy of patch relationships by combining authors’ qualitative analysis with developer interviews. Specifically, we first derived an initial set of relationship patterns by examining vulnerability descriptions, commit messages, code changes, etc. We then invited 20 developers in top-tier Internet companies with at least seven years of experience in open-source development, security maintenance, or vulnerability fixing, to validate and refine the taxonomy. Specifically, we provided them with randomly-sampled multi-patch CVE cases and the candidate taxonomy, and asked whether the relationship types were understandable, practically meaningful for downstream patch adoption, and whether any recurring types were missing. Their feedback was used to refine the category boundaries and descriptions. This process resulted in six recurring and actionable relationship types: \emph{Merge}, \emph{Mirror}, \emph{Better Solution}, \emph{Fixing-of-Fixing}, \emph{Collaboration}, and \emph{Separation}. This taxonomy is intended to cover dominant practical patterns. Rare or highly context-specific relationships may exist beyond these types, but they do not affect the main findings of this work.}

\revised{The above taxonomy construction follows common practice in empirical software engineering, where candidate concepts are first derived from empirical observations and then refined through human feedback. For example, Šmite et al. constructed a GSE (global software engineering) taxonomy by combining literature-based candidate terms with an expert survey, and showed that such empirically based taxonomies can support later research and decision-making~\cite{vsmite2014empirically}. Similarly, Doğan and Tüzün first manually collected code review bad practices from white and gray literature, then conducted interviews with experienced practitioners to refine the concepts, resulting in a taxonomy of code review smells that further supports tool development~\cite{dougan2022towards}. Beyond these, other studies from top-tier software engineering venues have also relied on qualitative analysis and practitioner-grounded evidence to derive concepts or taxonomies, such as References~\cite{sedano2017software, stol2016grounded}.}

\revised{For each patch relationship type, we analyze a representative real-world CVE example. Due to space limitations, the following subsections only give concise introductions, the detailed CVE pages, patch descriptions, code changes, and relationship analyses of these examples are available in our repository~\cite{detailedexample}}.

\subsection{``Merge'' Relationship}
\label{subsect:merge}

The \emph{Merge} relationship describes a scenario in which vulnerability patches developed on different branches are subsequently integrated into a single branch through a branch merge. In this case, one patch consolidates the effects of multiple related patches that exist on separate development lines, resulting in a unified fix for the vulnerability. At the Git level, such a merge patch is sometimes identifiable by having more than one parent commit, and its code content is formed by integrating the changes introduced by patches on the merged branches into the target branch’s code history.

As defined in Git's official documentation~\cite{chacon2014pro} and previous studies~\cite{10.1145/2568225.2568260,michaud2016recovering,10555642,10.1007/s10664-021-09951-x, zou2019branch}, this mechanism is widely used to consolidate parallel vulnerability-fixing efforts. 
For downstream maintainers, recognizing this relationship helps identify the consolidated fix and avoid separately reasoning about each merged patch when the merge patch is applicable, reducing the manual effort required during patch adoption. We analyze a real-world example of this relationship (CVE-2014-1829~\cite{CVE-2014-1829}, which reports a vulnerability in \emph{python-requests} allowing remote servers to obtain a \emph{netrc} password) in our supplementary document.

\subsection{``Mirror'' Relationship}
\label{subsect:mirror}

The \emph{Mirror} relationship describes a scenario in which identical or functionally equivalent vulnerability patches are independently applied to different versions (e.g., \emph{v}1.0 vs. \emph{v}2.0) or branches (e.g., \emph{main} vs. \emph{feature}) of the same codebase. Although these patches address the same vulnerability, they belong to independent code evolution paths with unique commit histories and parent pointers.

Mirror patches typically exhibit a high degree of similarity in their code changes. Minor differences may exist due to contextual factors, such as variations in surrounding code or line numbers across branches, but the underlying fixing logic remains equivalent. A typical mechanism that gives rise to this relationship is cherry-picking~\cite{7843626,10555642}, where a maintainer explicitly replicates a patch from one branch to another without merging the full branch history. For example, suppose a developer fixes a vulnerability with Patch \emph{P1} in a long-running \emph{feature} branch. Since this branch is not yet ready for full merging but the fix is critical for the stable \emph{main} branch, a maintainer uses ``\emph{git cherry-pick P1}'' to replicate the patch. This operation creates a new commit \emph{P2} in \emph{main} that carries the exact code changes from \emph{P1}. In this case, \emph{P1} and \emph{P2} form a \emph{Mirror} relationship. From the perspective of downstream maintainers, understanding this relationship helps them select the patch that matches the specific branch or version they depend on, without expending additional effort comparing multiple patches that are functionally equivalent. We analyze a real-world example of this relationship (CVE-2011-2931~\cite{CVE-2011-2931}, which reports a cross-site scripting vulnerability in \emph{Ruby on Rails} allowing attackers to inject arbitrary web scripts or HTML) in our supplementary document).

\subsection{``Better Solution'' Relationship}
\label{subsect:bettersolution}

The \emph{Better Solution} relationship describes a sequence of patches where a subsequent patch (can be denoted as \emph{P2}) refines a prior patch (can be denoted as \emph{P1}) by improving the implementation of the fix while preserving the same remediation mechanism for the vulnerability. Specifically, \emph{P2} focuses exclusively on improvements unrelated to the vulnerability fix itself, such as readability, maintainability, or performance (e.g., clarifying code comments, streamlining non-critical logic, or optimizing runtime efficiency), it does not alter the fundamental mechanisms that address the vulnerability (e.g., input validation rules, access control checks, or encryption implementations). \emph{P2} is typically dependent on \emph{P1}: it builds upon code structures, control logic, or functions introduced by \emph{P1}. While \emph{P2} may partially or even fully subsume \emph{P1}’s implementation at the code level, it does not replace the original security responsibility, which remains anchored in \emph{P1}.

This relationship commonly arises during post-fix refinement, where an initial security patch is subsequently polished for long-term maintenance or efficiency. From the perspective of downstream maintainers, the initial patch is usually sufficient to eliminate the vulnerability and can be applied promptly in time-sensitive scenarios, while later patches may be considered optionally to improve code quality or performance based on maintenance needs. We analyze a real-world example of this relationship (CVE-2014-1830~\cite{CVE-2014-1830}, which reports a vulnerability allowing remote servers to obtain sensitive user credentials) in our supplementary document.

\subsection{``Fixing-of-Fixing'' Relationship}
\label{subsect:fixingoffixing}

The \emph{Fixing-of-Fixing} relationship describes a scenario in which a subsequent patch corrects flaws that are unintentionally introduced by an earlier patch. In this relationship, the initial patch aims to remediate the original vulnerability but introduces new vulnerabilities or functional bugs due to incomplete logic, overlooked edge cases, implementation errors, and so on. The later patch directly targets the newly introduced problem and relies on the code structures or logic established by the earlier patch to perform the correction.

\emph{Fixing-of-Fixing} patches are inherently interdependent: applying only the initial patch may leave newly introduced vulnerabilities or bugs, while applying only the subsequent patch is ineffective without the foundational code it corrects. For downstream maintainers, these patches should be applied in order and typically expected to be included in the same release, since releasing them separately may lead to an unsafe intermediate state and thus expose users to new risks. We analyze a real-world example of this relationship (CVE-2019-12108~\cite{CVE-2019-12108}, which reports a denial-of-service vulnerability in \emph{MiniUPnP} caused by a null pointer dereference) in our supplementary document.

\subsection{``Collaboration'' Relationship}
\label{subsect:collaboration}

The \emph{Collaboration} relationship describes a scenario in which multiple patches jointly resolve a vulnerability, while none of them is sufficient on its own. In this relationship, an earlier patch introduces foundational security mechanisms (such as core data structures, variables, or preliminary checks) that are necessary but incomplete for eliminating the vulnerability. One or more subsequent patches then build on this foundation, extending or completing the remediation logic so that their combined effect fully addresses the vulnerability. This relationship commonly arises when fixing complex vulnerabilities that require multi-stage remediation, for example those involving layered systems, interdependent components, or intricate control flows.
For downstream maintainers, recognizing the collaboration relationship is important because it indicates that the patches are complementary and must be applied together to achieve a complete fix. We analyze a real-world example of this relationship (CVE-2020-25781~\cite{CVE-2020-25781}, which reports a permission bypass vulnerability allowing users without access to private issue notes to download attached files via direct URLs) in our supplementary document.

Note that while both \emph{Fixing-of-Fixing} and \emph{Collaboration} require sequential patch adoption, they differ in their intermediate states. \emph{Fixing-of-Fixing} introduces new flaws that make intermediate releases problematic, whereas \emph{Collaboration} results in incomplete but non-flawed intermediate fixes. Therefore, \emph{Collaboration} patches may be released separately under practical constraints, such as the tight schedule, the limitation of available manpower, and so on, as long as subsequent patches complete the remediation.

\subsection{``Separation'' Relationship}
\label{subsect:separation}

The \emph{Separation} relationship describes patches that address distinct aspects of a vulnerability or target the replicated vulnerable pattern across different locations within a codebase, without functional or code-level interdependence. In this relationship, patches operate on disjoint code regions, and no patch relies on code additions or modifications introduced by another. This relationship commonly arises in large-scale projects with parallel development activities, where different developers or teams independently address separate concerns, such as different modules, features, or flaw classes. As a result, the corresponding patches serve distinct purposes, evolve independently, and can be applied in any order or in parallel without affecting correctness or security. We analyze a real-world example of this relationship (CVE-2012-2673~\cite{CVE-2012-2673}, which reports integer overflow vulnerabilities in memory allocation functions that can lead to memory-related attacks) in our supplementary document.

\vspace{6pt}
\revised{The above six relationship types are derived from the qualitative analysis and developers survey, and are intended to capture recurring patch interaction patterns with clear downstream adoption implications, rather than to serve as an exhaustive taxonomy of all possible patch relationships.}

\section{Approach}
\label{sect:approach}

\subsection{Overview}
\label{subsect:overview}

\begin{figure*}[]
\centering
\includegraphics[width=0.9\textwidth]{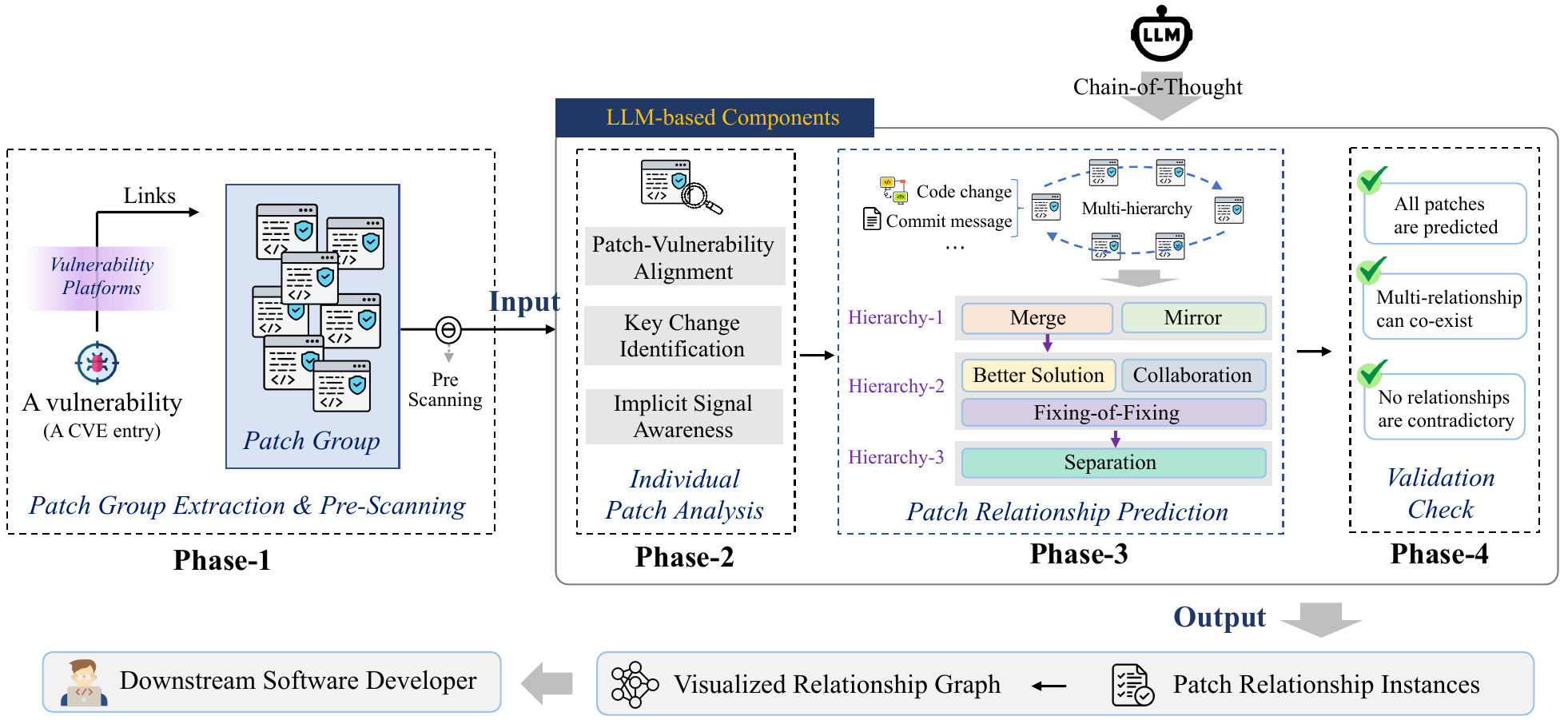} 
\caption{The overview of COMPASS}
\label{fig:overview}
\end{figure*}

The mission of COMPASS is to predict the relationships of multiple patches associated with a given vulnerability. COMPASS follows a four-phase pipeline for this goal. \textbf{In Phase-1} (\revised{Patch Group Extraction \& Pre-Scanning}), the method takes a vulnerability entry from public platforms (e.g., CVE, NVD, Snyk) as input. \revised{These platforms typically record links to the patches that fix the vulnerability}, allowing the required patch group to be obtained directly. \revised{After that, COMPASS determines some explicit patch relationships at an early stage via a pre-scanning step}. Then, the vulnerability (including the CVE ID and the description) and its patch group (including the commit ID, the branch/version information, the commit message, and the code change of each patch) will be fed into an LLM under the intuition of chain-of-thought. \textbf{In Phase-2} (Individual Patch Analysis), we ask the LLM to double-check \revised{the patch group provided by the vulnerability platform} for further determining whether each patch directly fixes the vulnerability. And we also ask the LLM to identify key components and files modified, and note any implicit signals in patch information, to facilitate the following analysis. \textbf{In Phase-3} (Patch Relationship Prediction), we send the description of the six types of patch relationships to the LLM, organize them into three hierarchies, and prompt the LLM to reason about relationships. \textbf{In Phase-4} (Validation Check), the LLM is asked to revisit its previous predictions, ensuring that all patches have been predicted and verifying that there are no contradictory relationships. Finally, COMPASS integrates the reasoning results of the above thinking processes and organize the output according to the predefined schema. The overview of COMPASS is given in Figure~\ref{fig:overview}.

\subsection{Phase-1: Patch Group Extraction \& Pre-Scanning}
\label{subsect:multiplepatchidentification}

Taking as input a vulnerability entry from public vulnerability platforms (e.g., CVE, NVD, Snyk), Phase-1 aims to identify its patch group (which serves as the input for subsequent analysis) and directly determine explicit patch relationships at an early stage. 

\revised{Vulnerability entries in public platforms typically provide explicit links to patches, allowing COMPASS to directly obtain the patch group.} 
Received the patch group, COMPASS performs a lightweight pre-scanning step over commit messages to identify explicit and high-confidence relationship evidence. For example, commit messages may contain indicative terms such as ``\emph{merge}'', ``\emph{merged in}'', or ``\emph{pull request}'', which strongly suggest the \emph{Merge} relationship, or terms such as ``\emph{refactor}'', ``\emph{optimize}'', and ``\emph{improve performance}'', which often signal the \emph{Better Solution} relationship~\cite{chen2018speedoo, rebai2020recommending, batoun2023empirical}. Such indicators serve as explicit textual evidence about the intent of a patch and, when present, allow the corresponding relationships to be determined with high confidence at an early stage.
\revised{Importantly, this pre-scanning step is only applicable when such explicit cues are present, which occurs in rare cases in real-world open-source projects~\cite{10.1145/3634737.3657007,reis2023secomlintlintersecuritycommit}. Prior studies on security commit messages have shown that such messages are often insufficiently informative. For example, Abreu et al. reported that 56.7\% of security commit messages were poorly documented~\cite{reis2023security}, and a large-scale replication study further confirmed that security commit messages remain not informative enough for security-focused tasks~\cite{islam2026informativeness}. That is, many vulnerability patches lack explicit relationship cues in their commit messages, motivating to reason over code changes together with message semantics.} For these majority cases, COMPASS proceeds with the LLM-based analysis pipeline (Phase-2 to Phase-4) to infer patch relationships.

Note that the aforementioned ``direct relationship determination'' in pre-scanning is performed at the relationship instance level. A relationship instance denotes a piece of relationship within a patch group, involving one or more patches. The resolved instances are not forwarded to the subsequent phases of COMPASS. As a reminder, each individual patch in such resolved instances is retained in the patch group, as it may also be involved in other unresolved relationship instances.

\subsection{Phase-2: Individual Patch Analysis}
\label{subsect:patchanalysis}

Phase-2 performs individual patch analysis as the preparation for relationship prediction. The vulnerability information (CVE ID and textual description) and all commits in the patch group (commit IDs and messages, branch/version information, and code changes) are fed into the LLM together with the prompt. Specifically, the prompt instructs the LLM to focus on three complementary aspects for each patch:

\begin{itemize}

    \item \revised{\textbf{Patch-Vulnerability Alignment.} Checking whether each patch in the patch group directly contributes to the remediation of the target vulnerability. This helps determine dependency-oriented relationships such as \emph{Better Solution}, \emph{Fixing-of-Fixing}, and \emph{Collaboration}, because they are often characterized by a sequential fixing chain, where later patches may refine, correct, or build on prior patches.}

    \item \textbf{Key Change Identification.} Identifying the key components, files, or functions modified by the patch. Knowing where changes occur provides evidence for detecting overlaps, dependencies, or individuality of patches, which could support distinguishing relationships of patches.

    \item \textbf{Implicit Signal Awareness.} Focusing on implicit cues available in commit messages and branch/version information (e.g., refinement-oriented wording or version-specific context). These cues are not sufficient for direct relationship determination on their own in the pre-scanning step, but can provide complementary signals when combined with code change analysis in later phases.
    
\end{itemize}

The goal of Phase-2 is not to directly predict or output patch relationships. It serves as a structured analysis scaffold, guiding the LLM to inspect each patch in isolation and extract key signals that are essential for the subsequent relationship reasoning. The per-patch observations produced in this phase are retained as internal context and are used as supporting evidence in Phase-3. By separating per-patch inspection from relationship reasoning, COMPASS encourages the LLM to first build a structured understanding of each patch, thereby reducing premature or inconsistent judgments when analyzing inter-patch relationships.

\subsection{Phase-3: Patch Relationship Prediction}
\label{subsect:patchrelationshipprediction}

Phase-3 predicts the relationships of patches by prompting the LLM to reason about unresolved relationship instances (i.e., instances that were not directly determined in the pre-scanning step) and identify typical patch relationship patterns, guided by the observations in Section~\ref{sect:typicalrelationshipsofpatches}. \revised{The prompt used in Phase-3 is given in our repository~\cite{phase3prompt}} due to space limitations. A relationship instance denotes a piece of relationship within a patch group, involving one or more patches. For relationship types \emph{Merge}, \emph{Mirror}, \emph{Better Solution}, \emph{Fixing-of-Fixing}, and \emph{Collaboration}, COMPASS records relationship instances in a dictionary form: $R_{ID\text{-}Type} = \{p_{ID\text{-}i}: [p_{ID\text{-}j}, ...], ...\}$, where $ID$ denotes the CVE ID, $Type$ denotes the name of one of these five relationship types, and $p_{ID\text{-}i}$, $p_{ID\text{-}j}$, ... denote the patches of the vulnerability. The dictionary's key represents the subject patch of the instance, and the value lists the corresponding object patches related to it, following the output schema described in the prompt (note that for \emph{Mirror}, if no patch naturally corresponds to a main branch or a primary maintained version, any patch in the instance can be chosen as the dictionary key). Since patches in \emph{Separation} do not exhibit code-level dependency and no subject patch exists, each such patch forms a relationship instance on its own, i.e., $R_{ID\text{-}``Separation"} = \{p_{ID\text{-}i}\}$.

COMPASS organizes patch relationship prediction into three hierarchies to guide the LLM through a structured reasoning process. Hierarchy-1 focuses on relationships that can be inferred from clear structural composition or high code-level similarity, Hierarchy-2 performs dependency-based reasoning, and Hierarchy-3 handles cases where no dependency is identified within the involved patch set.

\subsubsection{Hierarchy-1 (Merge and Mirror)} COMPASS first asks the LLM to identify \emph{Merge} and \emph{Mirror} relationships, as these two types often exhibit comparatively clear characteristics in code changes and branch/version contexts. For \emph{Merge}, the LLM checks whether a patch consolidates changes from multiple related patches on separate development lines, and distinguishes the merge patch from the merged patches, allowing downstream maintainers to easily identify the consolidated fix. For \emph{Mirror}, the LLM checks whether patches implement identical or highly similar fixes across different software versions or branches, with high similarity in both code changes and commit messages.

\subsubsection{Hierarchy-2 (Better Solution, Fixing-of-Fixing, Collaboration)} COMPASS then prompts the LLM to reason about dependency within the involved patch set, \revised{(e.g., whether a later patch is based on the earlier patch’s code, uses newly added or modified functions or variables, or whether the application order affects fixing effectiveness), which is explicitly described in the prompt}. Based on the identified dependency characteristics and the semantic intent inferred from code changes and commit messages, the LLM further analyzes each relationship instance into \emph{Better Solution}, \emph{Fixing-of-Fixing}, or \emph{Collaboration}. These three relationships reflect different dependency semantics, including refinement without security impact, correction of flaws introduced by earlier fixes, or complementary contributions toward a complete fix.

\subsubsection{Hierarchy-3 (Separation)} For relationship instances where the LLM does not identify dependency within the involved patch set, COMPASS guides the LLM to further examine the nature of their contributions. Specifically, the LLM reasons whether the patches address different aspects of the vulnerability independently, or perform fixes at multiple code locations where the vulnerable pattern is replicated. When such evidence is observed, the relationship instance is considered to exhibit \emph{Separation}.

To further enhance the reasoning, we augment the Phase-3 prompt with one-shot demonstrations: for each of the six relationship types, one real-world multi-patch CVE example (not included in our benchmark) is provided as an illustration of the expected inference and output format.

\subsection{Phase-4: Validation Check}
\label{subsect:validationcheck}

Phase-4 performs a validation check over the inferred relationship instances. This phase prompts the LLM to review the inferred results and ensure their completeness and consistency with respect to both the input patch group and the hierarchy-specific constraints.

First, the validation step ensures patch coverage: every patch in the input patch group should participate in at least one relationship instance, \revised{except for patches that have been identified as not contributing to the vulnerability fixing}. Second, the validation enforces hierarchy-aware labeling constraints. Relationship instances in Hierarchy-1 (\emph{Merge} and \emph{Mirror}) should be single-labeled, while instances in Hierarchy-2 (\emph{Better Solution}, \emph{Fixing-of-Fixing}, and \emph{Collaboration}) may carry multiple labels, as a single patch sometimes plays multiple roles within the same instance at the commit granularity. For Hierarchy-3 (\emph{Separation}), each instance is single-labeled, as it represents non-dependent patches and does not involve overlapping roles.

The final output for a CVE is the set of all validated relationship instances, which can be visualized as a relationship graph to facilitate downstream understanding and patch adoption.

\subsection{Running Example}
\label{subsect:runningexample}

We illustrate the running of COMPASS using a real-world example CVE-2013-5029~\cite{CVE-2013-5029} (here we only give a concise introduction due to space constraints, the complete version can be found in our repository~\cite{runningexample}). This CVE records a \emph{clickjacking}-related vulnerability in \emph{phpMyAdmin}. The NVD entry provides four fixing commits (\emph{240b833}, \emph{24d0eb5}, \emph{66fe475}, and \emph{da4042f}), which COMPASS directly collects as the patch group in Phase-1 (no explicit relationship can be determined by pre-scanning commit messages). In Phase-2, COMPASS inspects each patch individually and flags Patch \emph{66fe475} as a non-remediation commit because it only updates \emph{ChangeLog} without changing remediation logic. In Phase-3, COMPASS infers two relationship instances among the remaining remediation patches:

\vspace{3pt}
\begin{center}
\begin{minipage}{0.98\columnwidth}
\begin{tcolorbox}[
  width=\linewidth,
  colback=black!5,
  colframe=black,
  boxrule=0.5pt,
  arc=1mm,
  left=0.5mm,right=0.5mm,top=1mm,bottom=1mm,
  boxsep=0mm
]
\centering
\small
\setlength{\parskip}{0pt}
\setlength{\abovedisplayskip}{1pt}
\setlength{\belowdisplayskip}{1pt}
$R_{CVE\text{-}2013\text{-}5029\text{-}``Collaboration"}$: \{\emph{240b833}: [\emph{24d0eb5}]\},

\vspace{5pt}
$R_{CVE\text{-}2013\text{-}5029\text{-}``Fixing\text{-}of\text{-}Fixing"}$: \{\emph{24d0eb5}: [\emph{da4042f}]\}.
\end{tcolorbox}
\end{minipage}
\end{center}
\vspace{5pt}
This is because Patch \emph{240b833} introduces initial protection and related configuration semantics,  Patch \emph{24d0eb5} is written on top of it and adds an additional hiding strategy. This strategy introduces a flaw since the hiding behavior is applied too broadly, which may block legitimate page rendering. Patch \emph{da4042f} conditions the injected ``\emph{html}{\emph{display}: \emph{none};}'' to avoid unintended blocking.

\begin{figure}[]
\centering
\includegraphics[width=0.45\textwidth]{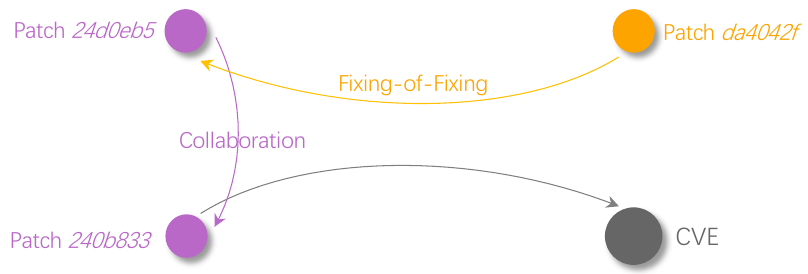} 
\caption{The relationships of patches of CVE-2013-5029}
\label{fig:runningexample}
\end{figure}

Finally, Phase-4 validates that all remediation patches participate in at least one relationship instance and that no inferred relationships are contradictory. The final output is also visualized as a graph, as shown in Figure~\ref{fig:runningexample}.

\section{Experimental Setup}
\label{sect:experimentalsetup}

\subsection{Research Questions (RQs)}
\label{subsect:researchquestions}

\subsubsection{RQ1: How effective are different LLMs when running COMPASS}
\label{subsubsect:rq1}

COMPASS uses LLM's reasoning capability in its essential part (Phase-2 to Phase-4). We run four representative models, \emph{DeepSeek-v3}~\cite{liu2024deepseek}, \emph{Claude-3.7}~\cite{claude37_2025}, \emph{Grok-3}~\cite{Grok-3}, and \emph{Gemini-2.5}~\cite{comanici2025gemini}, in the running of COMPASS, and compare their effectiveness.

\subsubsection{RQ2: \revised{How does COMPASS perform compared with learning-based and LLM-only baselines}}
\label{subsubsect:rq2}

\revised{To the best of our knowledge, there is no existing technique specifically designed for predicting relationships of multiple vulnerability patches. We therefore compare COMPASS with two categories of baselines: learning-based baselines built on pre-trained models widely used in related patch and commit understanding tasks, and an LLM-only baseline that directly prompts the best-performing LLM in RQ1 without COMPASS’s structured pipeline (details of the baselines are provided in Section~\ref{subsect:baselines}).}

\subsubsection{RQ3: How do key designs affect COMPASS’s effectiveness}
\label{subsubsect:rq3}

We ablate the one-shot prompting strategy used in Phase-3 and thus obtain $C_{zero\text{-}shot}$, to quantify the gain brought by it. Besides, since Phase-2 and Phase-4 perform preparatory analysis and post-hoc validation, respectively, we construct a variant $C_{pred\text{-}only}$ that removes these two phases, to examine whether they provide measurable benefits beyond the core phase (Phase-3).

\subsection{Baselines}
\label{subsect:baselines}

\subsubsection{Learning-based baselines}
\revised{We select widely-used pre-trained models such as \emph{CodeReviewer}~\cite{li2022automating}, \emph{CodeT5}~\cite{wang2021codet5}, and \emph{BERT}~\cite{devlin2019bert}. Although these models are not originally designed for patch relationship prediction, they have been successfully used in closely related tasks, including vulnerability patch localization~\cite{zhang2024dual, song2025not, li2024patchfinder} and issue-commit link recovery~\cite{zheng2025selink, zhang2023ealink, deng2024mtlink}. This comparison examines whether existing pre-trained models are sufficient for our task.}

We implement learning-based baselines by fine-tuning pre-trained models and cast the task as supervised classification over patch pairs. For a vulnerability description $Desc$, and a patch pair with commit information $Msg_1$, $Msg_2$ (including commit IDs, messages, and branch/version information) and code changes $CG_1$, $CG_2$, we construct the model input as $T$ = [[CLS], $Desc$, [SEP], $Msg_1$, $CG_1$, [SEP], $Msg_2$, $CG_2$, [SEP]], which is tokenized by the backbone tokenizer with padding or truncation to meet the model’s length limit. We also attach a lightweight classification head to map the backbone output to a 6-dimensional vector corresponding to the six relationship types studied in this work. As a patch pair may exhibit multiple relationship labels at the commit granularity, we treat the task as multi-label prediction using thresholding on each dimension (thresholds tuned on the validation set). This yields three baselines, $CR$, $C\text{-}T5$, and $BERT$ \revised{(fine-tuned from CodeReviewer, CodeT5, and BERT, respectively)}.

We additionally build a soft-prompting variant for each backbone for a more robust comparison. Soft prompting prepends a high-level task instruction to the model input, which can better align a general-purpose backbone with the target prediction task~\cite{wang2022no, wang2023prompt, zhang2024dual, li2024knowledge} and thus potentially strengthen the baseline. We prepend a task instruction $Soft$ (i.e., the instruction-only prompt extracted from COMPASS by removing all vulnerability and patch-specific fields, such as the CVE description and commit contents) to the model input. As such, the model input is $T_{soft}$ = [[CLS], $Soft$, [SEP], $Desc$, [SEP], $Msg_1$, $CG_1$, [SEP], $Msg_2$, $CG_2$, [SEP]]. This yields three baseline variants,  $CR\text{-}s$, $C\text{-}T5\text{-}s$, and $BERT\text{-}s$.

We evaluate \revised{learning-based} baselines using 10-fold cross-validation. In each fold, we split data into train, validation, and test sets with an 8:1:1 ratio (ensuring that test sets across folds are disjoint), and then aggregate the test results across folds. We train models with learning rate 1$e^{\text{-}4}$, batch size 48, up to 20 epochs, and the focal loss function.

\subsubsection{LLM-only baseline}
\revised{This baseline evaluates whether the effectiveness gain comes merely from the underlying LLM, or from the COMPASS pipeline that guides the LLM to reason about patch relationships in a structured manner. Since RQ1 shows that DeepSeek-v3 achieves the best result, we use DeepSeek-v3 as the representative LLM and directly provide it with the vulnerability information, patch information, and descriptions of the six relationship types, but remove COMPASS’s structured pipeline, including individual patch analysis, hierarchy-guided relationship reasoning, and validation checking. This yields a baseline $DS\text{-}only$}.

\subsection{Datasets}
\label{subsect:datasets}

\revised{We construct a benchmark dataset of 300 multi-patch CVEs randomly selected from CVE~\cite{cve-website}, NVD~\cite{nvd-website}, and Snyk~\cite{snyk-website}, spanning 2009$\sim$2024. This benchmark is completely disjoint from the 1,050 entries used for taxonomy construction in Section~\ref{sect:typicalrelationshipsofpatches}. The benchmark CVEs cover 213 open-source repositories, including projects such as \emph{apache\_tomcat}, \emph{phpmyadmin}, and \emph{linux}. In terms of popularity, 70.0\% of these repositories have more than 1K stars, including 20.7\% with more than 10K stars and 31.0\% with 3K$\sim$10K stars. Meanwhile, 39.9\% have more than 1K forks, including 7.0\% with more than 10K forks and 8.9\% with 3K$\sim$10K forks. These statistics suggest that the benchmark involves many widely-used open-source projects.}

\revised{The dataset contains 694 patches, where each CVE has 2$\sim$7 patches. For annotation, two experts having experience in software engineering and security independently labeled relationship instances, based on the vulnerability description and each patch’s information such as commit messages and code changes, strictly following guidelines aligned with the relationship descriptions in Section~\ref{sect:typicalrelationshipsofpatches}. The Cohen’s Kappa is 87.98\%, indicating strong agreement. Disagreements were resolved by involving a third expert and reaching consensus through discussion. In total, we annotate 473 relationship instances, which are available at the online querying website: \urlstyle{rm}{\color{black}\url{https://patch-relation.com}.}}

\subsection{Metrics}
\label{subsect:metrics}

We evaluate COMPASS using Precision, Recall, and F1-score, classic effectiveness metrics for prediction tasks~\cite{sokolova2009systematic, li2019deepfl, li2021fault, yang2024learning}, and report results for each relationship type \revised{(reporting metrics per relationship type mitigates the impact of potential minority classes)}. The metrics are computed by matching predicted relationship instances against manually annotated instances. For each relationship type $Type$, we collect: (\emph{i}) the set of predicted relationship instances of this type across all CVEs, denoted by $\mathbb{R}_{Type}$, (\emph{ii}) the set of manually-annotated relationship instances for this type, denoted by $\mathbb{G}_{Type}$. We then derive the set of correct predictions $\mathbb{C}_{Type}$ $\subseteq$ $\mathbb{R}_{Type}$, where an instance is counted as correct only if it matches an annotated instance under the same type. Specifically, we match predicted and annotated relationship instances in a type-aware manner. For \emph{Merge}, \emph{Better Solution}, \emph{Fixing-of-Fixing}, and \emph{Collaboration}, an instance matches only if the dictionary’s key-value pairs are identical. For \emph{Mirror}, if there is no designated subject patch (i.e., any patch can serve as the dictionary key), we ignore the key choice and match instances by their patch sets. For \emph{Separation}, each patch alone forms an instance, so a match reduces to equality of $p_{ID\text{-}i}$. Given the above symbols, we calculate $Recall_{Type}$ as $|\mathbb{C}_{Type}| / |\mathbb{G}_{Type}|$, $Precision_{Type}$ as $|\mathbb{C}_{Type}| / |\mathbb{R}_{Type}|$, and $F1\text{-}Score_{Type}$ as the harmonic mean of $Recall_{Type}$ and $Precision_{Type}$.

\section{Result and Analysis}
\label{sect:resultandanalysis}

\subsection{RQ1: Effectiveness of different LLMs}
\label{subsect:rq1}

We report the per-type Recall, Precision, and F1-Score of COMPASS when instantiated with four LLMs listed in Section~\ref{subsubsect:rq1} in Table~\ref{tbl:rq1}. Overall, DeepSeek-v3 achieves promising effectiveness across relationship types, yielding higher F1-score for all six relationships (\emph{Merge} 80.49\%, \emph{Mirror} 87.36\%, \emph{Better Solution} 62.50\%, \emph{Fixing-of-Fixing} 69.23\%, \emph{Collaboration} 63.64\%, and \emph{Separation} 82.86\%).

A closer look shows that different LLMs can be best on Precision for different relationship types. Specifically, Gemini-2.5 achieves the highest Precision on \emph{Merge} (88.89\%), Claude-3.7 is the most precise on \emph{Fixing-of-Fixing} (78.95\%). These results suggest that some models tend to make more conservative predictions for certain relationship types, leading to fewer false positives. But such precision advantages do not necessarily translate into the best overall effectiveness when Recall is considered jointly.

DeepSeek-v3 attains the highest Recall and F1-Score across all relationship types under the same COMPASS pipeline, indicating a better balance between coverage and correctness in instance-level inference. Notably, the effectiveness gap (e.g., F1-Score) is most visible on the dependency-driven relationships handled in Hierarchy-2, including \emph{Better Solution}, \emph{Fixing-of-Fixing}, and \emph{Collaboration}, where COMPASS explicitly guides the model to hierarchically reason about inter-patch dependency rather than relying on shallow cues. 

These results suggest that COMPASS’s structured pipeline enables patch relationship inference across different LLM backbones, where the LLM mainly serves as a replaceable component. Among the evaluated models, DeepSeek-v3 yields better overall effectiveness, therefore, we adopt DeepSeek-v3 as the default LLM for COMPASS in the remaining RQs.

\begin{table}[t]
\centering
\scriptsize
\setlength{\tabcolsep}{2.5pt}
\renewcommand{\arraystretch}{1}

\caption{Different LLMs' reasoning capabilities}
\begin{adjustbox}{max width=\columnwidth}
\begin{tabular}{llcccc}
\toprule
 \;\;\;\;\textbf{\scriptsize{Metric}} & \textbf{\scriptsize{Type}} & \makecell{\;\textbf{\scriptsize{DeepSeek-v3}}} & \makecell{\;\textbf{\scriptsize{Claude-3.7}}} & \makecell{\;\textbf{\scriptsize{\;\;\;Grok-3\;\;\;}}} & \makecell{\textbf{\scriptsize{Gemini-2.5}}} \\
\midrule
\multirow{6}{*}{\textbf{Recall\bm{$_\text{\tiny{\emph{Type}}}$}}}    & Merge            & \textbf{94.29\%}    & 65.71\%    & 80.00\% & 68.57\%    \\
                           & Mirror           & \textbf{87.02\%}     & 84.73\%    & 80.92\% & 76.34\%    \\
                           & Better  & \textbf{83.33\%}    & 50.00\%     & 50.00\% & \textbf{83.33\%}    \\
                           & F-o-F & \textbf{72.00\%}      & 60.00\%    & 56.00\% & 64.00\%    \\
                           & Colla    & \textbf{73.68\%}      & 57.89\%    & 52.63\% & 47.37\%    \\
                           & Sepa       & \textbf{76.10\%}      & 49.80\%     & 51.39\% & 60.56\%    \\ \hline
\multirow{6}{*}{\textbf{Precision\bm{$_\text{\tiny{\emph{Type}}}$}}\;\;\;\;\;} & Merge            & 70.21\%      & 85.19\%    & 65.12\% & \textbf{88.89\%}    \\
                           & Mirror          & \textbf{87.69\%}    & 81.02\%     & 79.10\% & 76.34\%    \\
                           & Better  & \textbf{50.00\%}       & 23.08\%    & 20.69\% & 40.00\%    \\
                           & F-o-F & 66.67\%      & \textbf{78.95\%}     & 58.33\% & 64.00\%    \\
                           & Colla    & \textbf{56.00\%}       & 18.03\%   & 26.32\% & 19.15\%    \\
                           & Sepa       & \textbf{90.95\%}       & 90.58\%    & 84.87\% & 84.92\%    \\ \hline
\multirow{6}{*}{\textbf{F1-Score\bm{$_\text{\tiny{\emph{Type}}}$}}}  & Merge            & \textbf{80.49\%}        & 74.19\%   & 71.79\% & 77.42\%    \\
                           & Mirror            & \textbf{87.36\%}     & 82.84\%  & 80.00\% & 76.34\%    \\
                           & Better  & \textbf{62.50\%}      & 31.58\%    & 29.27\% & 54.05\%    \\
                           & F-o-F & \textbf{69.23\%}       & 68.18\%     & 57.14\% & 64.00\%    \\
                           & Colla    & \textbf{63.64\%}     & 27.50\%    & 35.09\% & 27.27\%    \\
                           & Sepa       & \textbf{82.86\%}     & 64.27\%     & 64.02\% & 70.70\%    \\ \hline
\end{tabular}
\end{adjustbox}

\vspace{2mm}
\tiny{Note: Due to space limitations, ``Better'', ``F-o-F'', ``Colla'', and ``Sepa'' denote ``Better Solution'', ``Fixing-of-Fixing'', ``Collaboration'', and ``Separation'', respectively. The same abbreviations are used in the following tables.}

\label{tbl:rq1}
\end{table}

\subsection{RQ2: Comparison with baseline techniques}
\label{subsect:rq2}

\begin{figure}[]
\centering
\includegraphics[width=0.5\textwidth]{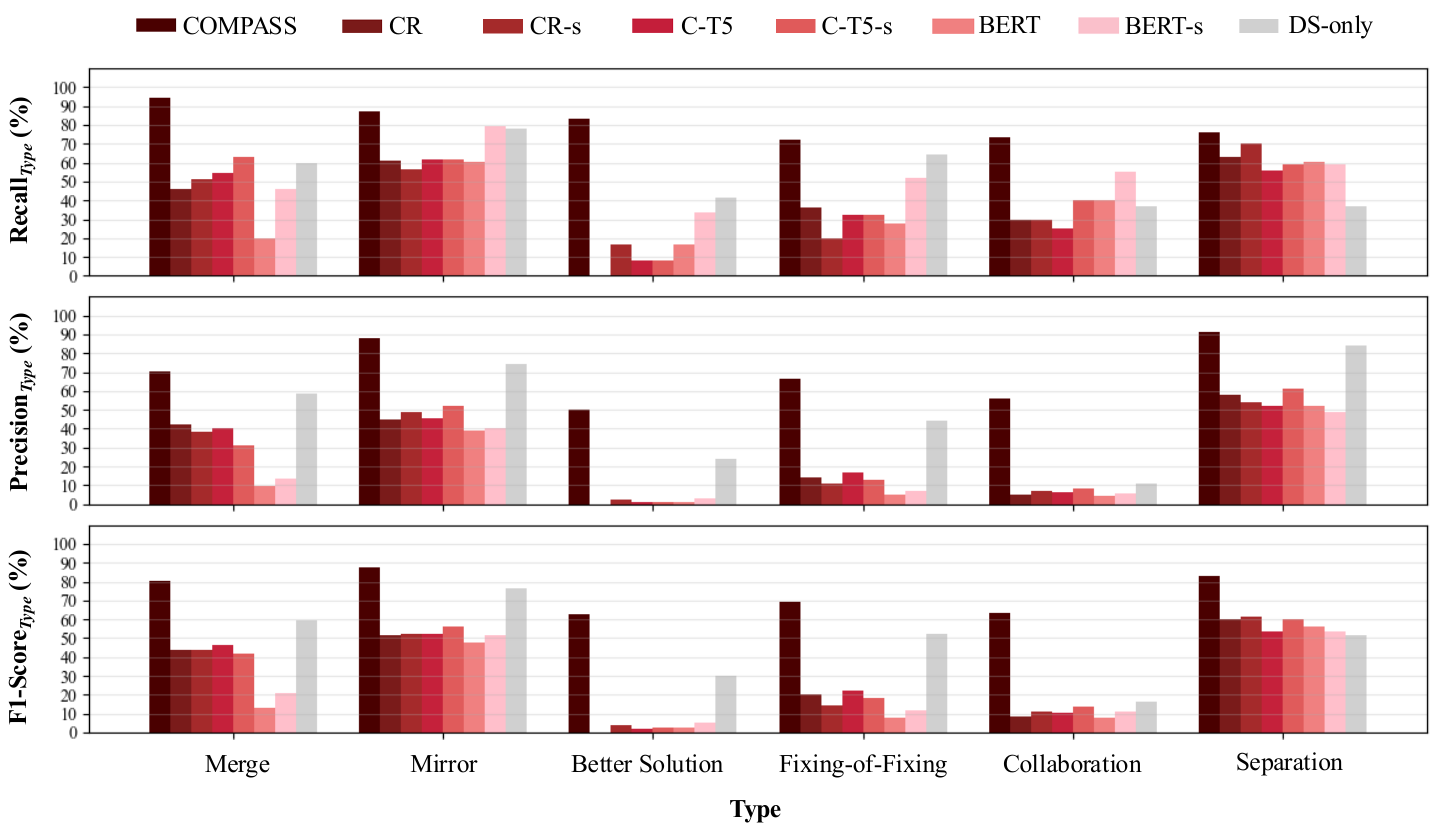} 
\caption{Comparison between COMPASS and baselines}
\label{fig:rq2}
\end{figure}

\revised{We report the comparison between COMPASS and learning-based baselines $CR$, $C\text{-}T5$, $BERT$, $CR\text{-s}$, $C\text{-}T5\text{-}s$, and $BERT\text{-}s$, and LLM-only baseline $DS\text{-}only$ in Table~\ref{tbl:rq2} and Figure~\ref{fig:rq2}. Across all six relationship types, COMPASS achieves the best results in terms of Recall, Precision, and F1-Score. More specifically, for each relationship type, we use the best baseline on that type as the reference and calculate COMPASS’s relative F1-Score improvement, averaging the improvements over all types gives an overall improvement of 85.04\%. This demonstrates high competitiveness over pairwise fine-tuning of generic pre-trained models, and shows that the gain of COMPASS does not merely come from the LLM itself, but also from its structured pipeline ($DS\text{-}only$ uses the same strongest LLM but is consistently worse than COMPASS).}

COMPASS's advantage is particularly evident on the dependency-driven relationships in Hierarchy-2. For \emph{Better Solution}, \emph{Fixing-of-Fixing}, and \emph{Collaboration}, COMPASS attains F1-Scores of 62.50\%, 69.23\%, and 63.64\%, respectively, \revised{greatly outperforming learning-based and LLM-only baselines. This gap suggests that instance-level dependency reasoning, which is explicitly guided by COMPASS’s hierarchical inference design, is difficult to recover through pairwise classification or LLM reasoning alone. For relationships whose evidence is more distinctive in code changes or version/branch context (rather than explicit message cues), the baselines exhibit improved effectiveness but still remain notably behind COMPASS. For example, on \emph{Merge} and \emph{Mirror}, the best baselines achieve F1-Scores of 59.17\% and 76.12\%, respectively, compared to 80.49\% and 87.36\% for COMPASS, respectively. And on \emph{Separation}, the best baseline reaches 61.22\%, while COMPASS achieves 82.86\%.}

Among the learning-based baselines, we further analyze the behavior of the soft-prompt variants, whose gains are limited and occasional. Notable improvements can be observed in multiple cases, for example, \emph{BERT-s} increases \emph{Fixing-of-Fixing} Recall from 28.00\% (\emph{BERT}) to 52.00\%, and also improves \emph{Mirror} Recall from 60.31\% (\emph{BERT}) to 79.23\%. In contrast, some cases show little change, e.g., the F1-Score of \emph{Merge} for \emph{CR-s} is from 43.84\% (\emph{CR}) to 43.90\%, and others even degrade, e.g., the Precision of \emph{Fixing-of-Fixing} for \emph{C-T5-s} drops from 17.02\% (\emph{C-T5}) to 13.11\%. Importantly, the soft-prompt variants remain weak on dependency-driven relationships (Hierarchy-2) overall: even the best soft-prompt result reaches F1-Scores only 5.30\% on Better Solution (\emph{BERT-s}), 18.60\% on \emph{Fixing-of-Fixing} (\emph{C-T5-s}), and 13.56\% on \emph{Collaboration} (\emph{C-T5-s}), far below COMPASS.

\begin{table}[t]
\centering
\huge
\setlength{\tabcolsep}{2.5pt}
\renewcommand{\arraystretch}{1.12}

\caption{\revised{Comparison between COMPASS and baselines}}
\begin{adjustbox}{max width=\columnwidth}
\begin{tabular}{llcccccccc}
\toprule
 \;\;\;\;\textbf{Metric} & \textbf{\;Type\;\;} & \makecell{\LARGE{\textbf{COMPASS}}} &\makecell{\;\;\;\;\textbf{CR}\;\;\;\;\;} & \makecell{\;\;\;\textbf{CR\text{-}s}\;\;\;} & \makecell{\;\;\;\textbf{C\text{-}T5}\;\;\;} & \makecell{\;\;\textbf{C\text{-}T5\text{-}s}\;\;} & \makecell{\;\;\textbf{BERT}\;\;\;} & \makecell{\;\textbf{BERT\text{-}s}\;\;} & \makecell{\;\textbf{DS\text{-}only}} \\
\midrule
\multirow{6}{*}{\textbf{Recall\bm{$_{Type}$}}}    & Merge            & \textbf{94.29\%}     & 45.71\%  & 51.43\%    & 54.29\%    & 62.86\% & 20.00\% & 45.71\%  & 60.00\%  \\
                           & Mirror           & \textbf{87.02\%}     & 61.07\%  & 56.49\%    & 61.83\%    & 61.83\% & 60.31\% & 79.23\%  & 77.86\% \\
                           & Better  & \textbf{83.33\%}     & 00.00\%  & 16.67\%    & 8.33\%    & 8.33\% & 16.67\% & 33.33\%  &  41.67\%  \\
                           & F-o-F & \textbf{72.00\%}     & 36.00\%  & 20.00\%    & 32.00\%    & 32.00\% & 28.00\% & 52.00\%  & 64.00\% \\
                           & Colla    & \textbf{73.68\%}     & 30.00\%  & 30.00\%    & 25.00\%    & 40.00\% & 40.00\% & 55.00\%  &  36.84\% \\
                           & Sepa       & \textbf{76.10\%}     & 63.20\%  & 70.40\%    & 55.60\%    & 59.20\% & 60.40\% & 59.20\%  &  37.05\% \\ \hline
\multirow{6}{*}{\textbf{Precision\bm{$_{Type}$}}\;} & Merge            & \textbf{70.21\%}     & 42.11\%  & 38.30\%    & 40.43\%    & 31.43\% & 9.72\% & 13.79\%  & 58.33\%\\
                           & Mirror          & \textbf{87.69\%}     & 44.69\%  & 48.68\%    & 45.25\%    & 51.92\% & 39.11\% & 40.53\%  & 74.45\% \\
                           & Better  & \textbf{50.00\%}     & 00.00\%  & 2.11\%    & 1.18\%    & 1.37\% & 1.41\% & 2.88\%  & 23.81\% \\
                           & F-o-F & \textbf{66.67\%}     & 14.29\%  & 11.11\%    & 17.02\%    & 13.11\% & 4.76\% & 6.74\%  & 44.44\% \\
                           & Colla    & \textbf{56.00\%}     & 5.04\%  & 6.82\%     & 6.58\%    & 8.16\% & 4.49\% & 5.98\%  &  10.61\% \\
                           & Sepa       & \textbf{90.95\%}     & 57.66\%  & 54.15\%    & 52.06\%    & 61.41\% & 52.25\% & 49.01\%   & 83.78\% \\ \hline
\multirow{6}{*}{\textbf{F1-Score\bm{$_{Type}$}}}  & Merge            & \textbf{80.49\%}     & 43.84\%  & 43.90\%    & 46.34\%    & 41.90\% & 13.08\% & 21.19\%  & 59.17\%  \\
                           & Mirror            & \textbf{87.36\%}     & 51.61\%  & 52.30\%    & 52.26\%    & 56.45\% & 47.45\% & 51.40\%   & 76.12\% \\
                           & Better  & \textbf{62.50\%}     & 00.00\%  & 3.74\%    & 2.06\%    & 2.35\% & 2.60\% & 5.30\%  & 30.30\% \\
                           & F-o-F & \textbf{69.23\%}     & 20.45\%  & 14.29\%    & 22.22\%    & 18.60\% & 8.14\% & 11.93\%  &  52.48\% \\
                           & Colla    & \textbf{63.64\%}     & 8.63\%  & 11.11\%     & 10.42\%    & 13.56\% & 8.08\% & 10.78\%  & 16.49\% \\
                           & Sepa       & \textbf{82.86\%}     & 60.31\%  & 61.22\%    & 53.77\%    & 60.29\% & 56.03\% & 53.62\%   &  51.39\% \\ \hline
\end{tabular}
\end{adjustbox}

\label{tbl:rq2}
\end{table}

\subsection{RQ3: Ablation of key designs}
\label{subsect:rq3}

We report the ablation results on two key designs of COMPASS: removing the one-shot demonstrations in Phase-3 ($C_{zero\text{-}shot}$) and using only the prediction stage ($C_{pred\text{-}only}$, i.e., without the supporting phases), \revised{as shown in Table~\ref{tbl:rq3}}. Overall, both variants degrade effectiveness across relationship types, indicating that the one-shot strategy and the full pipeline design contribute positively to instance-level relationship inference. Specifically, removing one-shot demonstrations leads to consistent F1-Score drops, with the largest degradations on the dependency-driven relationships in Hierarchy-2, e.g., \emph{Better Solution} (from 62.50\% to 52.94\%), \emph{Fixing-of-Fixing} (from 69.23\% to 59.64\%), and especially \emph{Collaboration} (from 63.64\% to 32.14\%). On the other hand, removing the supporting phases also causes effectiveness losses on several types, notably \emph{Collaboration} (from 63.64\% to 36.11\%) and \emph{Separation} (from 82.86\% to 51.08\%), suggesting that Phase-2 and Phase-4 provide useful preparatory patch understanding and post-hoc validation for the final predictions.

We note two exceptions where $C_{zero\text{-}shot}$ is slightly higher (Recall$_{Mirror}$: from 87.02\% to 87.79\%, and Precision$_{Separation}$: from 90.95\% to 91.22\%). These improvements are accompanied by F1-Scores decrease, suggesting that the gains come from a more aggressive or conservative bias (higher Recall or Precision) rather than better overall instance inference.

\begin{table}[t]
\centering
\scriptsize
\setlength{\tabcolsep}{2.5pt}
\renewcommand{\arraystretch}{1.05}

\caption{Ablation of key designs in COMPASS's pipeline}
\begin{adjustbox}{max width=0.8\columnwidth}
\begin{tabular}{llccc}

\toprule
\textbf{Metric} & \textbf{Type}\;\;\;\; & \textbf{\;\;\;\;COMPASS\;\;\;\;} & \bm{$C_{zero\text{-}shot}$}\;\;\;\;  & \;\;\bm{$C_{pred\text{-}only}$} \\[-0.5ex]
\midrule

\multirow{6}{*}{\textbf{Recall\bm{$_\text{\tiny{\emph{Type}}}$}}}    &  Merge           & \textbf{94.29\%}     & 85.71\%  & 82.86\%      \\
                           & Mirror          & 87.02\%     & \textbf{87.79\%}  & 85.50\%      \\
                           & Better  & \textbf{83.33\%}     & 75.00\%  & 75.00\%   \\
                           &  F-o-F & \textbf{72.00\%}     & 68.00\%  & 68.00\%      \\
                           & Colla   & \textbf{73.68\%}     & 47.37\%  & 68.42\%    \\
                           & Sepa      & \textbf{76.10\%}     & 53.78\%  & 37.85\%     \\ \hline
\multirow{6}{*}{\textbf{Precision\bm{$_\text{\tiny{\emph{Type}}}$}}}\;\;\;\;\;\; &  Merge            & \textbf{70.21\%}     & 69.77\%  & 69.05\%     \\
                           &  Mirror         & \textbf{87.69\%}     & 80.42\%  & 84.85\%       \\
                           & Better & \textbf{50.00\%}     & 40.91\%  & 29.03\%      \\
                           & F-o-F & \textbf{66.67\%}     & 53.12\%  & 53.12\%     \\
                           & Colla   & \textbf{56.00\%}     & 24.32\%  & 24.53\%      \\
                           & Sepa    & 90.95\%     & \textbf{91.22\%}  & 78.51\%    \\ \hline
\multirow{6}{*}{\textbf{F1-Score\bm{$_\text{\tiny{\emph{Type}}}$}}}  &  Merge            & \textbf{80.49\%}     & 76.92\%  & 75.32\%      \\
                           & Mirror            & \textbf{87.36\%}     & 83.94\%  & 85.17\%     \\
                           &  Better  & \textbf{62.50\%}     & 52.94\%  & 41.86\%    \\
                           &  F-o-F & \textbf{69.23\%}     & 59.64\%  & 59.65\%      \\
                           & Colla  & \textbf{63.64\%}     & 32.14\%  & 36.11\%       \\
                           & Sepa    & \textbf{82.86\%}     & 67.67\%  & 51.08\%       \\ \hline
\end{tabular}
\end{adjustbox}

\label{tbl:rq3}
\end{table}

\section{Discussion}
\label{sect:discussion}

\begin{figure}[]
\centering
\includegraphics[width=0.42\textwidth]{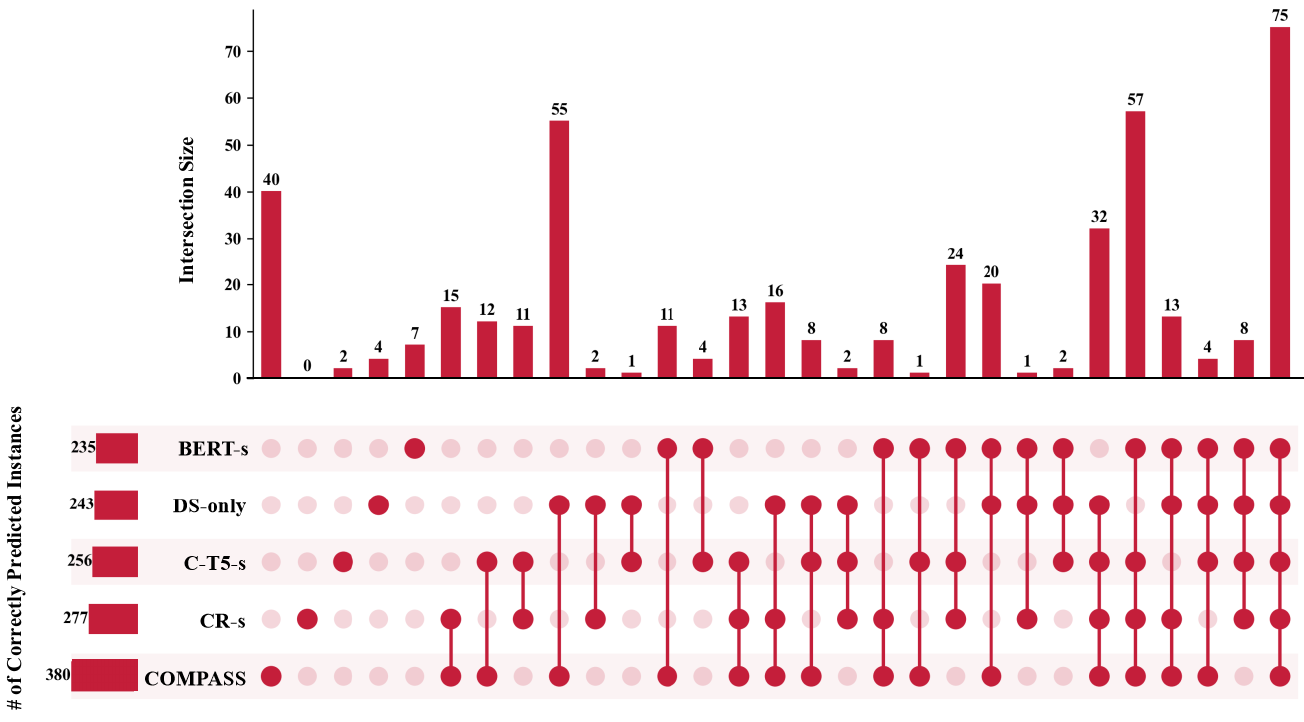} 
\caption{Unique Correctly Predicted Relationship Instances}
\label{fig:discussion}
\end{figure}

Beyond the three standard metrics, we further examine the uniqueness of correct predictions over COMPASS and four \revised{representative baselines (\emph{DS-only}, \emph{CR-s}, \emph{C-T5-s}, and \emph{BERT-s})}, as shown in an UpSet plot~\cite{lex2014upset} in Figure~\ref{fig:discussion}. The horizontal bars on the left report each method’s total number of correctly predicted instances, while each column in the dot matrix encodes a specific intersection pattern (filled dots indicate the methods included in that intersection), whose size is given by the corresponding bar on top.

Overall, COMPASS recovers the largest set of annotated instances (380), outperforming \revised{selected baselines (235$\sim$277)}. More importantly, COMPASS also contributes the largest unique portion: \revised{40} instances are correctly recovered only by COMPASS, suggesting that it adds substantial additional coverage beyond overall effectiveness advantages. Meanwhile, the UpSet plot contains multiple non-empty intersections excluding COMPASS, indicating that baselines occasionally recover instances that COMPASS misses. These baseline-only cases are small in absolute number compared with COMPASS’s unique portion, but they show that different modeling paradigms can complement each other on certain corner cases.

\section{Threats to Validity}
\label{sect:threatstovalidity}

\revised{Our study is subject to several threats to validity. First, the taxonomy and manual labels may involve subjective judgment. To mitigate this, we constructed the taxonomy with practitioner feedback (following well-established taxonomy construction practices in empirical software engineering), used a benchmark disjoint from the taxonomy construction set, and annotated each patch group by experts independently, achieving a Cohen’s Kappa of 87.98\%. Second, data leakage is a potential risk for LLM-based software engineering techniques. In our study, this risk is extremely slim because the task-critical ground truth is the patch relationship label, rather than publicly available CVE or patch information. Such relationship labels were newly created by us and are not available in public vulnerability platforms or existing datasets. Moreover, $DS\text{-}only$ performs worse than COMPASS, suggesting that the gain does not merely come from memorization. Third, effectiveness may vary across ecosystems or projects. Our benchmark randomly samples 300 CVEs from CVE, NVD, and Snyk, spans 2009$\sim$2024, and covers 213 repositories, but larger-scale and industrial validation is still needed.}

\section{Related Work}
\label{sect:relatedwork}

Prior studies have shown that multi-patch vulnerabilities are prevalent, but most of them focus on locating relevant fixes or characterizing the phenomenon, rather than inferring how patches \revised{of the same CVE} relate to each other. Li et al.~\cite {li2017large} and Tan et al.~\cite {tan2021locating} conduct empirical studies on multi-patch vulnerability remediation and summarize common characteristics and scenarios. Hommersom et al. study automated identification and ranking of fix commits, and note that a vulnerability may correspond to multiple fixing commits in practice~\cite{hommersom2024automated}. Woo et al. further investigate patch effectiveness and show that a non-trivial portion of disclosed fixes are incomplete and may require supplementary patches, which increases the complexity of downstream remediation~\cite{woo2025large}. Most relevant to our study, Xu et al. examine the quality of CVE–patch mappings and track patches across branches or repositories, highlighting that a CVE does not always map to a single patch and that equivalent fixes may appear in different development lines~\cite{xu2022tracking}. In contrast, \revised{COMPASS goes beyond identifying or characterizing multiple patches: it infers instance-level relationships within a patch group and outputs structured relationship instances that can directly guide downstream patch adoption.}

\section{Conclusion}
\label{sect:conclusion}

This paper studies the practically important yet under-explored problem of understanding relationships of multiple patches for the same vulnerability, which directly affects downstream patch adoption. \revised{We propose COMPASS, an automated LLM-based approach that predicts instance-level patch relationships under six recurring and actionable relationship types derived from qualitative analysis and developers validation. COMPASS outperforms both compact learning-based baselines and an LLM-only baseline on a benchmark of 300 multi-patch CVEs, the average improvement reaches 85.04\%. We also release the benchmark and an online querying website to support community reuse.}

\revised{In future work, we plan to explore broader relationship patterns and evaluate COMPASS in larger-scale and industrial downstream patch adoption scenarios.}

\bibliographystyle{IEEEtran}
\balance
\bibliography{software}

\end{document}